# Anomalous Circular Polarization of Magneto-Photoluminescence from Individual CdSe Nanocrystals

H. Htoon[1], S. A. Crooker[2], M. Furis[2‡], S. Jeong[1†], Al. L. Efros[3], V. I. Klimov[1*]

[1]*Chemistry Division and the Center for Integrated Nanotechnologies, Los Alamos National Laboratory, Los Alamos, NM 87545*
[2]*National High Magnetic Field Laboratory, Los Alamos, NM 87545*
[3]*Naval Research Laboratory, Washington, DC 20375*

We study the low-temperature magneto-photoluminescence (PL) from individual CdSe nanocrystals. Nanocrystals having a small "bright" exciton fine structure splitting ($\Delta_{XY} < 0.5$ meV) exhibit a conventional left- and right-circularly polarized Zeeman PL doublet in applied magnetic fields. In contrast, nanocrystals with large $\Delta_{XY}$ (>1 meV) show an anomalous magneto-PL polarization, wherein the lower-energy peak becomes circularly polarized with increasing field, while the higher-energy peak remains linearly polarized. This unusual behavior arises from strong mixing between the absorbing and emitting bright exciton levels due to strong anisotropic exchange interactions.

75.50.Pp, 78.55.Et, 71.55.Gs, 78.66.Hf, 78.67.D

[*] *Corresponding author: klimov@lanl.gov*
[‡] *Current Address: Department of Physics, University of Vermont, Burlington, VT 05405*
[†] *Current Address: Korea Institute of Machinery and Materials, Daejon, 305-343, KOREA*

A Zeeman splitting between otherwise degenerate spin eigenstates under applied magnetic fields is a fundamental quantum mechanical effect that has been studied in many different quantum systems ranging from atoms and molecules to electron-hole excitations (excitons) in both bulk [1] and nanostructured [2-4] semiconductors. Recently, a significant research effort has focused on field-dependent spin phenomena in semiconductor quantum dots (QDs) [2, 3, 5, 6], motivated by a desire to understand the fundamental spin structure of QD electronic states and by potential technological applications in solid-state spintronics [7] and quantum information processing [8].

Zeeman effects in self-assembled QDs grown by epitaxial methods have been extensively studied in both QD ensembles [6, 9] and in single QDs [2, 10] using magneto-photoluminescence (PL). In cylindrically symmetric QDs, a magnetic field $\boldsymbol{B}$ applied along the epitaxial growth direction ($\hat{z}$) splits the lowest-energy, spin-degenerate, optically-allowed ("bright") exciton into two states with net spin projections +1 and -1. In the Faraday geometry ($\boldsymbol{k} \parallel \boldsymbol{B}$), these states emit right- or left-circularly polarized light ($\sigma^{\pm}$) as depicted in Fig. 1(a). However, many epitaxial QDs possess a significant asymmetry in the X-Y plane, normal to the growth direction $\hat{z}$. In these QDs, the long-range (anisotropic) part of the electron-hole exchange interaction mixes the $|\pm 1\rangle$ bright states to produce a "fine structure" of two *linearly* and *orthogonally* polarized dipoles, $|X,Y\rangle = (|+1\rangle \pm |-1\rangle)/\sqrt{2}$ with energy splitting $\Delta_{XY}$ at $\boldsymbol{B}$=0 [6, 11,12]. With increasing $\boldsymbol{B}$, the PL polarization from these states evolves from linear to elliptical to left/right circular in the limit where the Zeeman energy $E_z = g_{ex} \mu_B B \gg \Delta_{XY}$ [Fig. 1(b)]. Here, $\mu_B$ is the Bohr magneton and $g_{ex}$ is the bright exciton g-factor. This "traditional" picture has been experimentally confirmed in numerous studies of epitaxially-grown QDs [2, 6, 13].

Previous experiments have also addressed Zeeman effects in semiconductor nanocrystal quantum dots (NQDs) synthesized by colloidal chemistry. However, until now, magneto-PL methods have been applied exclusively to ensembles of NQDs [14], where the PL linewidth (>50 meV) is inhomogeneously broadened by the NQD size distribution. As a result, Zeeman splittings between right- and left-circularly polarized PL are not well resolved. Recent studies of colloidal CdSe NQDs using PL line-narrowing methods in high *B* also revealed a linearly-polarized XY "fine-structure" splitting of bright excitons [15], similar to that observed in epitaxial dots. These results were supported by detailed single-NQD experiments in zero field [16], which demonstrated that $\Delta_{XY}$ scales with inverse NQD volume. Because of their small sub-5 nm radii, NQDs exhibit very large values of $\Delta_{XY}$ from 1-3 meV, which is larger than typical sub-meV values reported for epitaxial dots [10-13] (although exceptions do exist [17]).

In this Letter, we study polarized PL from individual CdSe NQDs at low temperatures and in magnetic fields to 5 T. The unusually strong anisotropic exchange splitting $\Delta_{XY}$ found in many NQDs dramatically modifies the magneto-PL polarization of the emitting "bright" exciton. We observe two distinctly different and systematic behaviors: In all NQDs exhibiting small $\Delta_{XY}$ (<0.5 meV) we observe a traditional $\sigma^{\pm}$-polarized Zeeman doublet, from which we measure, for the first time, the exciton *g*-factor of single nanocrystals. In contrast, all NQDs exhibiting large $\Delta_{XY}$ (>1 meV) show a highly anomalous PL polarization wherein the lower-energy PL peak becomes circularly polarized with increasing *B* while the higher energy peak *remains linearly polarized*. We show that strong mixing between the emitting and absorbing bright-exciton states, which results from anisotropic exchange interactions in these NQD, accounts for this effect.

In these experiments we use a single-dot magneto-PL system that allows for simultaneous detection of both orthonormal PL polarization components. This system largely mitigates the undesirable effects of PL line wandering (spectral diffusion) and intensity fluctuations (blinking) that are typical in single-NQD studies [18]. In this work, dilute dispersions of CdSe/ZnS core/shell NQDs in polymethylmethacrylate are deposited onto quartz substrates (< 0.5 nanocrystals/$\mu m^2$), and mounted in a 4 K optical cryostat in the room-temperature bore of a 5 T superconducting magnet. The NQDs are excited by a continuous wave 532 nm laser through a 40X objective (0.4 numerical aperture), and PL from individual NQDs is collected by the same objective. A quarter-wave plate then converts right- and left-circularly polarized PL to horizontal- and vertical-polarizations, which are then spatially separated by a Wollaston prism and directed through an imaging spectrometer onto the upper and lower halves of a cooled charge-coupled device. In this way, both $\sigma^+$ and $\sigma^-$ PL spectra are acquired simultaneously, greatly simplifying the quantitative analysis of magneto-PL data in the unavoidable presence of line wandering and blinking. To measure angle-resolved *linear* PL polarizations, the quarter-wave plate is replaced with a half-wave plate [16]. We studied hundreds of individual, randomly-oriented CdSe/ZnS NQDs emitting in the range 1.95 - 2.25 eV. We analyze the magneto-PL from those dots that are oriented with $\hat{c}$-axes nearly along (within ~30° of) the observation axis. For these 28 dots -- ~10% of the total -- the two bright exciton dipoles project nearly equally onto to observation plane, permitting straightforward quantitative analysis of magneto-polarization effects. [19]

Figure 1(c) shows polarization-resolved PL from a single NQD from **B**=0-5 T. Linear polarization analysis at 0 T indicates a small fine structure splitting $\Delta_{XY}$ = 0.28 mV (lowest spectrum), indicating weak anisotropic exchange [16]. The $\sigma^+$ and $\sigma^-$ PL components overlap at zero field, while increasing $B$ leads to a clear Zeeman splitting between these components that reaches $E_Z \sim 0.7$ meV at 5 T. Both PL peaks become nearly 100% circularly polarized.

The spectra in Fig. 1(c) exhibit relatively weak PL line wandering and intensity fluctuations. More typical single-NQD data are shown in Fig. 1(d), where scan-to-scan fluctuations are more apparent. At any given $B$, however, the relative $\sigma^+$ and $\sigma^-$ PL intensities as well as the Zeeman splitting remain unchanged. Again, both $\sigma^+$ and $\sigma^-$ PL peaks are clearly resolved at 5 T ($E_Z = 0.75$ meV) and are nearly 100% circularly polarized. Figures 1(e,f) show that, for both NQDs, $E_Z$ increases linearly with $B$, indicating "bright" exciton $g$-factors of 2.4 and 2.6 respectively, in approximate agreement with earlier Faraday rotation [20] and magneto-PL studies [6] of II-VI nanocrystal ensembles. Importantly, *all* NQDs that developed a well-resolved and $\sigma^{\pm}$-polarized Zeeman doublet also exhibited a small fine structure splitting $\Delta_{XY} < 0.5$ meV.

A strikingly different magneto-PL polarization is observed for NQDs having a large intrinsic $\Delta_{XY}$. PL from one such NQD is shown in Figs. 2(a). From linearly-polarized PL analysis [Fig. 2(a), the bottom spectrum], $\Delta_{XY} \sim 1.0$ meV ($B = 0$). Further, the fact that both linear components have nearly identical amplitudes indicates that this NQD's crystalline $\hat{c}$-axis is aligned nearly parallel to the observation axis (and to $B$) [16][19]. The top three spectra of Figure 2(a) track $\sigma^{\pm}$-resolved magneto-PL from this same NQD from 0 - 5 T. The lower-energy PL peak becomes progressively more circularly polarized with increasing $B$, as expected in the traditional picture of a Zeeman effect. However, the polarization evolution of the higher-energy peak is quite anomalous: It does not become circularly polarized but instead remains linearly polarized. The marked difference between these two PL peaks is particularly clear when comparing their field-dependent degree of circular polarization (DCP), defined as the normalized intensity ratio ($I_{\sigma+} - I_{\sigma-}$)/($I_{\sigma+} + I_{\sigma-}$) [Figure 2(b)]. We emphasize that a similarly anomalous magneto-PL polarization is observed from *all* individual NQDs that possess a large $\Delta_{XY}$ – spectra from another such NQD is shown in Fig. 2(c).

To understand this anomalous PL polarization in CdSe NCs, we calculate the band-edge exciton energies and oscillator strengths in the presence of both short-range (isotropic) and long-range (anisotropic) electron-hole exchange interactions. The short range exchange interaction, $H_e$, is proportional to the scalar product ($\sigma \cdot J$) of the electron Pauli spin-1/2 matrices $\sigma$ and the hole spin-3/2 matrices $J$ [21 22]. In wurtzite CdSe NQDs, this exchange interaction splits the band-edge exciton into five exciton levels labeled by their angular momentum projection $F$ on the crystalline *c*-axis [Fig. 3 (a)] [21, 23]. These include two states with $F = 0$ ($0^U$ and $0^L$ for the upper and lower levels, respectively), two two-fold degenerate "bright" exciton states with $F = \pm 1$ ($1^L$ and $1^U$), and the two-fold degenerate "dark" exciton state with $F = \pm 2$. The lowest-energy bright exciton, $1^L$, is responsible for PL, while the higher-lying $1^U$ bright exciton has larger oscillator strength and dominates the NQD absorption.

The anisotropic long-range exchange interaction, $H_{ae}$, is associated with the XY asymmetry of the NQD. $H_{ae}$ mixes the 5 exciton levels and lifts their angular momentum degeneracy. One can write: $H_{ae} = (\delta/4) (\sigma_+ J_+ + \sigma_- J_-)$, where $J_\pm = J_x \pm i J_y$, $\sigma_\pm = \sigma_x \pm i\sigma_y$, and $\delta$ is a phenomenolgical constant proportional to the non-analytic part of the electron-hole exchange integral. Following [20], in the basis of independent electron-hole pair states $|s_e, M_h\rangle$ ($s_e = \uparrow, \downarrow$ is the electron spin projection and $M_h = \pm 1/2, \pm 3/2$ is the hole angular momentum projection), $H_{ae}$ has the following nonzero elements: $\langle\uparrow,-1/2| \sigma_+J_+|\downarrow,-3/2\rangle = \langle\uparrow,+3/2|\sigma_+J_+|\downarrow,+1/2\rangle = -i\sqrt{3}\,\delta'$, $\langle\downarrow,-3/2| \sigma_-J_-|\uparrow,-1/2\rangle = \langle\downarrow,+1/2| \sigma_-J_-|\uparrow,+3/2\rangle = i\sqrt{3}\,\delta'$, and $\langle\downarrow,-1/2| \sigma_-J_-|\uparrow,1/2\rangle = -\langle\uparrow,1/2|\sigma_+J_+|\downarrow,-1/2\rangle = i2\delta'$, where $\delta' \sim \delta$.

For CdSe NQDs emitting at 2.0 eV, Fig. 3(a) shows the calculated energies of the four linearly-polarized exciton states deriving from $1^L$ and $1^U$ as a function of $\Delta_{XY}$. $\mathbf{H}_{ae}$ causes a much larger splitting of the $1^U$ bright exciton compared to that of the $1^L$ exciton, a consequence of its much larger oscillator strength. As a result, the upper component of the $1^L$ doublet and the lower component of the $1^U$ doublet (which we denote by their zero-field labels, $1^{L+}$ and $1^{U-}$) can approach each another, leading to significant state mixing in applied magnetic fields and to the observed anomalous magneto-PL polarization, as discussed below.

To compare with the data, we include the electron and hole interactions with magnetic field [20]: $\hat{H}_B = (1/2)\, g_e\mu_B\, (\boldsymbol{\sigma}\cdot\mathbf{B}) - g_h\mu_B\, (\boldsymbol{\kappa}\cdot\mathbf{B})$, where $\boldsymbol{\kappa}$ is the hole angular momentum and $g_e\,(g_h)$ is the g-factor of the 1S electron ($1S_{3/2}$ hole). Figures 3(b,c) show the calculated field-dependent exciton energies for CdSe NQDs having small and large $\Delta_{XY}$ (0.28 and 1.0 meV, respectively). We consider only the four states derived from the bright excitons ($1^{L\pm}$ and $1^{U\pm}$), as only these couple to circularly-polarized light in the Faraday geometry. Their $\mathbf{B}$-dependent transition probabilities and DCP are plotted in Figs. 3(d, e).

In NQDs having small $\Delta_{XY}$ [Fig. 3(d)], the DCP of the emitting $1^{L\pm}$ states quickly saturates at $\pm 1$ with increasing $\mathbf{B}$, as expected in the traditional picture of the Zeeman effect. This calculated behavior qualitatively reproduces the experimental data in Fig. 1. Figures 3(b, d) also show an interesting peculiarity expected at very high magnetic fields ($B \approx 40$ T) where $1^{L+}$ and $1^{U-}$ exciton states approach one another and anticross (marked by dotted circle). Interstate mixing at this point completely modifies the PL polarization, as shown in Fig. 3(d): The DCP for these states drops to zero at $B \approx 40$ T and then inverts sign at higher fields.

In contrast, the calculations demonstrate that NQDs having a large anisotropic exchange (large $\Delta_{XY}$) can be expected to show an anomalous PL polarization at low $B$ [Fig. 3(e)]. In these NQDs the $1^{L+}$ and $1^{U-}$ states are energetically close to each other at $B = 0$ and can therefore easily be mixed even under low applied fields [25]. As a result, PL from the $1^{L+}$ state *remains* nearly linearly polarized until the Zeeman energy $E_Z \gg \Delta_{XY}$ (at ~15 T), while emission from the $1^{L-}$ state becomes circularly polarized. This model of mixing induced by the large anisotropic exchange in colloidal NQDs agrees well with the experimental data (see Fig. 2).

In summary, we have performed polarized magneto-PL studies of individual colloidal NQDs. By analyzing both polarization- and spectrally-resolved PL in magnetic fields to 5 T, we measure exciton *g*-factors in individual CdSe NQDs. A traditional Zeeman effect, wherein both Zeeman-split exciton states emit circularly-polarized light, is observed for NQDs possessing a small degree of anisotropic exchange. In contrast, NQDs with large anisotropic exchange exhibit an anomalous PL polarization, wherein the higher-energy PL peak remains linearly polarized with increasing $B$ even though the lower-energy PL peak acquires a significant degree of circular polarization. This behavior, not reported in other types of semiconductor nanostructures (e.g., epitaxial QDs), results from the very large anisotropic exchange interactions (>1 meV) that are typical of ultrasmall colloidal NQDs, which leads to significant mixing between the emitting ($1^L$) bright exciton and the higher-lying absorbing ($1^U$) bright exciton. These effects are most clearly observed from those NQDs that are oriented along the observation direction (~10% of dots in a randomly-oriented sample). However, in shape-controlled NQDs that can self-align into larger coherent assemblies, these polarization effects may dominate the magneto-PL of ensemble samples. These effects can then be exploited for characterizing, for example, the degree of asymmetry via spectroscopic means, which would complement traditional microstructural studies.

We thank Anna Trugman for technical assistance. This work was supported by the DOE Office of Basic Energy Sciences and Los Alamos LDRD funds. V.I.K. and H.H. acknowledge partial support by the DOE Center for Integrated Nanotechnologies jointly operated by Los Alamos and Sandia National Laboratories. Al.L.E acknowledges the support of the ONR.

**Figure Captions**

FIG. 1. (Color online) (a) Zeeman splitting of bright $|\pm 1\rangle$ excitons in cylindrically symmetric QDs. (b) Zeeman splitting of a bright exciton in asymmetric QDs having an intrinsic linearly-polarized XY fine structure in zero B. (c) Circularly polarized ($\sigma^\pm$) PL (top three spectra) from a nearly symmetric NQD from 1-5 T, showing clear Zeeman splitting and strong circular polarization. Linearly-polarized PL at $B = 0$ (bottom spectrum) shows a small intrinsic fine structure splitting ($\Delta_{XY} = 0.28$ meV). T=4 K. (d) Same, from another NQD, which shows significant fluctuations in the PL line position. By simultaneously acquiring PL spectra for both $\sigma^+$ and $\sigma^-$ polarizations, we can still reliably measure the Zeeman splitting. (e,f) The energy splitting between $\sigma^\pm$ PL peaks (circles) and between linearly-polarized PL peaks (squares) from (c,d), respectively.

FIG. 2. (Color online) (a) Linearly-polarized PL at $B = 0$ from a single NQD, showing a large fine structure splitting ($\Delta_{XY} = 1.1$ meV) of the emitting ($1^L$) bright exciton (the bottom spectrum) along with spectra of circularly polarized PL for $B$ from 1, 3, and 5 T (three upper spectra). With increasing $B$, the low energy peak becomes $\sigma^+$ polarized while the higher-energy peak shows no circular polarization. (b) DCP of these two peaks (circles and squares, respectively). (c) The $\sigma^\pm$ PL at 5 T and the linearly polarized PL at 0 T from another NQD that exhibits a large $\Delta_{XY}$ and similarly anomalous polarization behavior.

FIG. 3. (Color online) (a) The four linearly-polarized bright exciton levels that derive from $1^L$ and $1^U$, as a function of $\Delta_{XY}$, which characterizes the strength of anisotropic exchange. While $1^{L,U}$ excitons are two-fold degenerate for cylindrically symmetric NQDs (left) they split as the NQD becomes asymmetric in the XY plane and $\Delta_{XY}$ increases (right). States $1^{L+}$ and $1^{U-}$ approach as $\Delta_{XY}$ increases. (b,c) Calculated $1^{L\pm}$ and $1^{U\pm}$ exciton energies versus $B$ ($\| c$) for NQDs having a small and large anisotropic exchange ($\Delta_{XY} = 0.28$ and 1.0 meV, respectively). The dashed circle shows the "anticrossing" region. (d,e) Calculated DCP for these states.

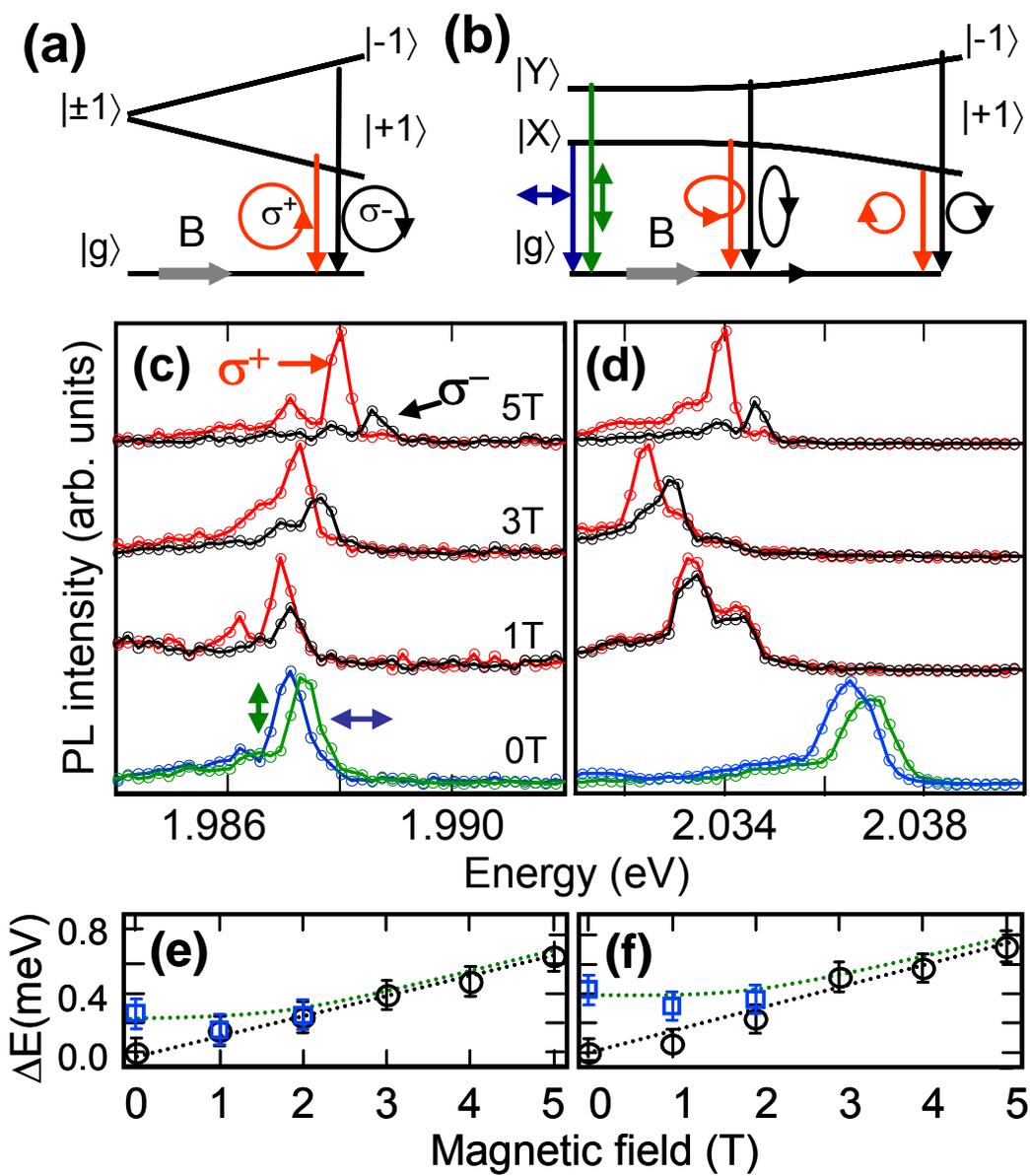

Figure 1, H. Htoon et al.,

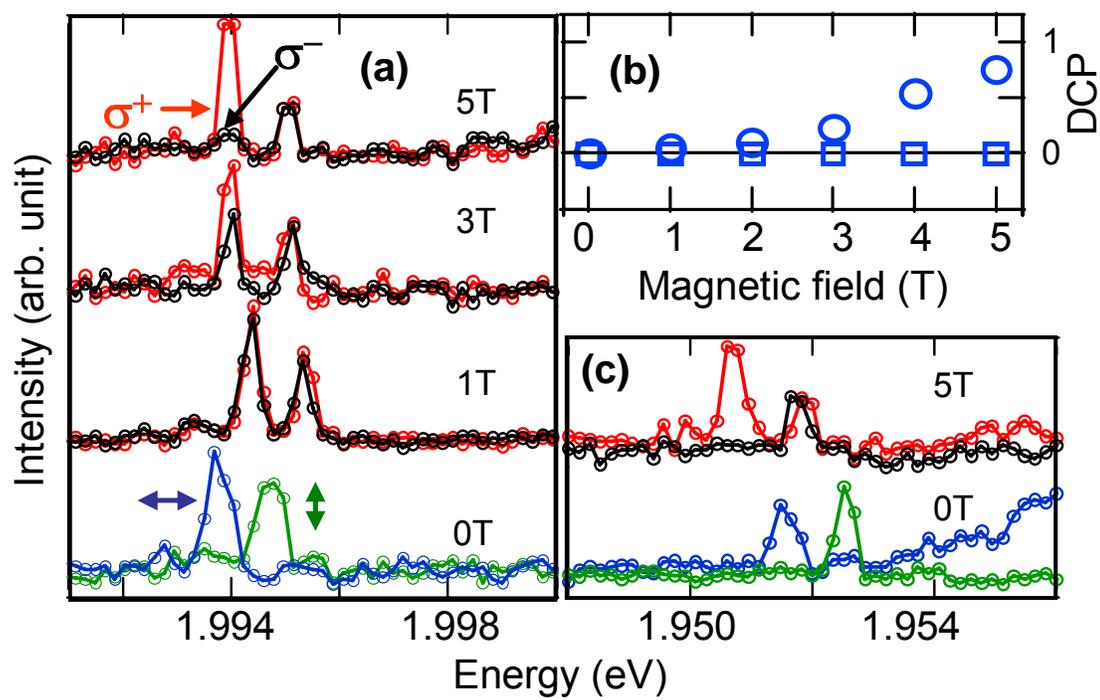

Figure 2, H. Htoon et al.,

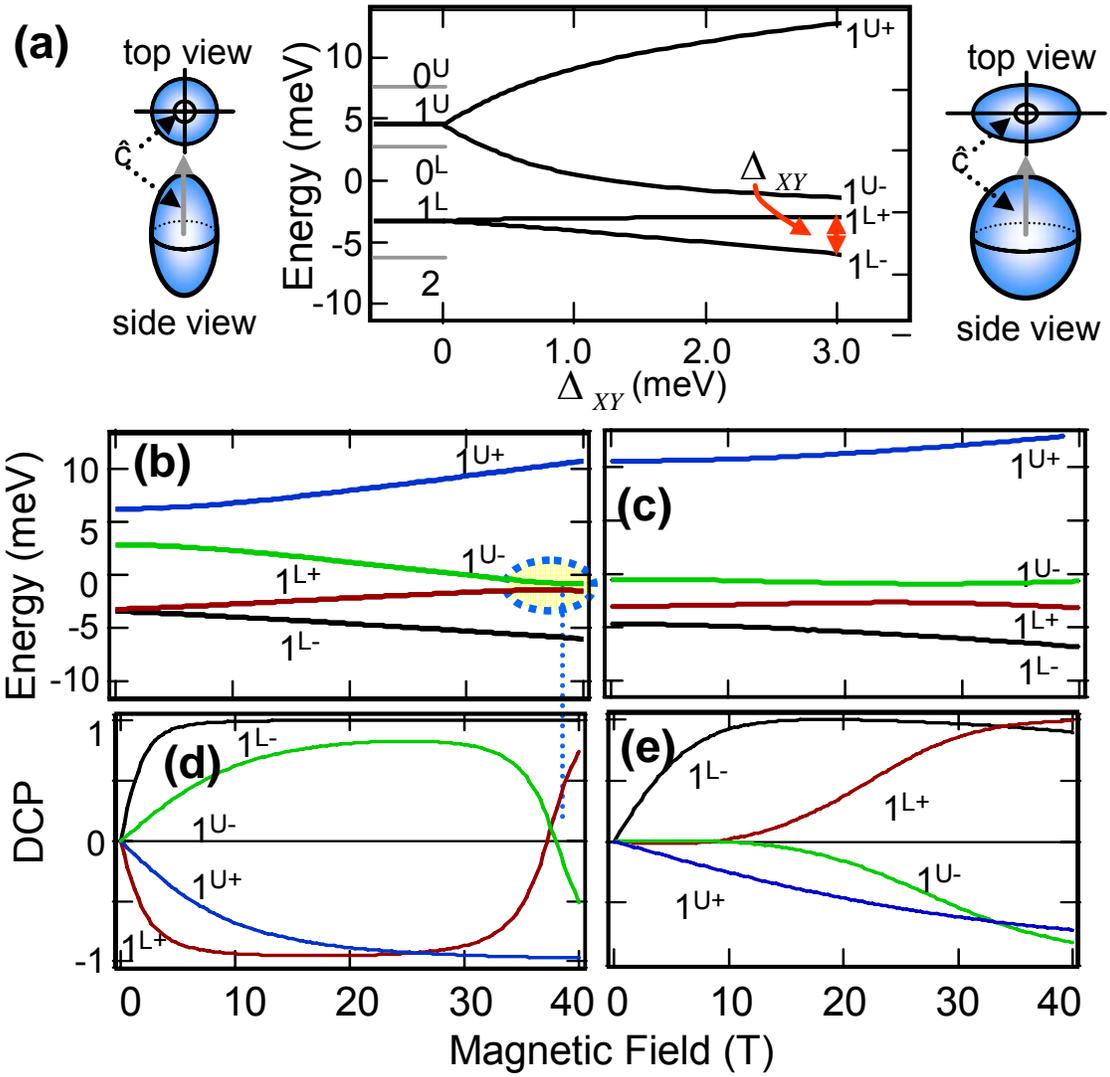

Figure 3, H. Htoon et al.,

15